# Ferromagnetic resonance in epitaxial films: Effects of lattice strains and voltage control via ferroelectric substrate

N. A. Pertsev,[1,2] H. Kohlstedt,[1] and R. Knöchel[3]

[1]*Nanoelektronik, Technische Fakultät, Christian-Albrechts-Universität zu Kiel, D-24143 Kiel, Germany*

[2]*A. F. Ioffe Physico-Technical Institute, Russian Academy of Sciences, 194021 St. Petersburg, Russia*

[3]*Hochfrequenztechnik, Technische Fakultät, Christian-Albrechts-Universität zu Kiel, D-24143 Kiel, Germany*

The phenomenon of ferromagnetic resonance (FMR) provides fundamental information on the physics of magnetic materials and lies at the heart of a variety of signal processing microwave devices. Here we demonstrate theoretically that substrate-induced lattice strains may change the FMR frequency of an epitaxial ferromagnetic film dramatically, leading to ultralow and ultrahigh resonance frequencies at room temperature. Remarkably, the FMR frequency varies with the epitaxial strain *nonmonotonically*, reaching minimum at a critical strain corresponding to the strain-induced spin reorientation transition. Furthermore, by coupling the ferromagnetic film to a ferroelectric substrate, it becomes possible to achieve an efficient *voltage control* of FMR parameters. In contrast to previous studies, we found that the tunability of FMR frequency varies with the applied electric field and strongly increases at critical field intensity. The revealed features open up wide opportunities for the development of advanced tunable magnetoelectric devices based on strained nanomagnets.

## I. INTRODUCTION

The ferromagnetic resonance (FMR), predicted by Landau and Lifshitz[1] in 1935 and observed experimentally by Griffiths[2] in 1946, arises from the precessional motion of the magnetization vector about the equilibrium direction that is driven by a microwave magnetic field. The studies of FMR in bulk crystals, thin films, and multilayers provided fundamental information on the gyromagnetic ratio, magnetic anisotropy, interlayer exchange coupling, and spin dynamics in magnetic materials.[3-5] The FMR also has a significant practical importance, being employed in signal processing microwave devices such as tunable band-stop filters, phase shifters, and signal-to-noise enhancers.[6-9]

The tuning of resonance condition is usually realized via the application of a static magnetic field to the ferromagnet. However this approach has the disadvantages of slow tuning speed and high energy consumption associated with magnetic field generation. On the other hand, the existence of a magnetoelastic contribution to the free energy indicates that the FMR can be also affected by lattice strains.[10-12] This effect was employed in ferrite-piezoelectric composites to tune the reso-nance magnetic field $H_r$ by an applied voltage,[13-15] which allows high tuning speed with low energy consumption. It was found that $H_r$ varies with voltage almost linearly, which was attributed to strains induced in the piezoelectric material and transmitted to the ferrite via the mechanical interfacial coupling. The experimental observations were described by a linear theory taking into account small piezoelectric deformations generated by applied field, which induce lattice strains in the ferromagnetic material.[14-16]

Although the linear approximation seems to be sufficient for the theoretical description of FMR in weakly strained bulk magnetic materials and composites, it is expected to give poor results for epitaxial ferromagnetic films and nanostructures. Indeed, considerable strains are usually generated in epitaxial heterostructures during growth, especially in films with *nanoscale* thicknesses, where strain relaxation via formation of misfit dislocations does not occur. Moreover, large voltage-induced strains of ~1% may be created in a thin ferromagnetic film attached to a ferroelectric substrate with ultrahigh piezoelectric coefficients, such as $Pb(Zn_{1/3}Nb_{2/3})O_3$–$PbTiO_3$ (PZN-PT) or $Pb(Mg_{1/3}Nb_{2/3})O_3$–$PbTiO_3$ (PMN-PT). As lattice strains can affect the magnetization direction and even induce spin reorientation transitions,[17-23] dramatic changes in the FMR are expected for epitaxial hetero-structures. This expectation is supported by the observation of a nonlinear electric-field-induced variation of the FMR frequency in the FeGaB/PZN-PT hybrid.[24]

In this paper, we describe the strain effect on FMR with the aid of *nonlinear* thermodynamic theory. Both the initial epitaxial strain imposed on a ferromagnetic film and additional variable strains produced by the substrate piezoelectric deformations under applied electric field are taken into account in the calculations of FMR frequency and resonance magnetic field.

## II. RESONANCE CONDITIONS IN STRAINED FILMS

Consider a homogeneously magnetized ferromagnetic film subjected to a static magnetic field **H** and a microwave magnetic field **h**($t$). The Landau-Lifshitz-Gilbert torque equation of motion for the dynamic magnetization $\mathbf{M}(t) = \mathbf{M}_s + \delta\mathbf{M}(t)$ may be written as

$$\frac{d\mathbf{M}}{dt} = -\gamma\mu_0(\mathbf{M}\times\mathbf{H}_{\text{eff}}) + \frac{G}{\gamma M^2}\left(\mathbf{M}\times\frac{d\mathbf{M}}{dt}\right), \qquad (1)$$

where $\gamma$ is the gyromagnetic ratio, $\mu_0$ is the permeability of vacuum, $\mathbf{H}_{\text{eff}}$ is the effective magnetic field, and $G$ is the Gilbert damping parameter.[1,5] Well below the Curie temperature, the magnitude $M$ of dynamic magnetization can be regarded as a fixed quantity.[11] Then Eq. (1) may be rewritten in terms of direction cosines $m_i$ ($i$ = 1,2,3) of the magnetization vector **M** and linearized with respect to small deviations $\delta m_i \ll 1$ from the static orientation $\{m_1^0, m_2^0, m_3^0\}$, which are induced by a weak microwave field with amplitude $|\mathbf{h}| \ll |\mathbf{H}|$. The effective magnetic field involved in Eq. (1) is defined by the relation $\mu_0\mathbf{H}_{\text{eff}} = -\partial F/\partial\mathbf{M}$, where $F$ is the Helmholtz free energy density of a ferromagnet. $\mathbf{H}_{\text{eff}}$ is the sum of static field **H**, microwave field **h**, demagnetizing field depending on the demagnetizing factors $N_i$ of a ferromagnetic sample, and additional terms resulting from the energy of magnetocrystalline anisotropy and magnetoelastic energy.

Let us calculate the effective field $\mathbf{H}_{\text{eff}}$ for an epitaxial film of a cubic ferromagnet with the (001) crystallographic orientation in the paramagnetic state. Restricting our analysis to the film thicknesses beyond a few-monolayer range, for which the surface contribution to the total energy[25] may be neglected, we can consider the Helmholtz free energy density $\Delta F$ of interior homogeneous magnetic state only. As epitaxial strains lower the symmetry of crystal lattice, the magnetic anisotropy of a strained ferromagnetic film differs from that of a bulk ferromagnet. This feature can be described by calculating the energy density $\Delta F$ with the account of magnetoelastic and strain energies and mechanical boundary conditions imposed on an epitaxial film.[22] Adding the Zeeman energy and the demagnetizing field energy to the expression derived for $\Delta F$ in Ref. 22 and using the reference frame ($x_1$, $x_2$, $x_3$) shown in Fig. 1, we obtain

$$\begin{aligned}\Delta F &= K_1 m_1^2 m_2^2 + \left(K_1 + \frac{B_1^2}{2c_{11}} - \frac{B_2^2}{2c_{44}}\right)(m_1^2 + m_2^2)m_3^2 \\ &+ K_2 m_1^2 m_2^2 m_3^2 + B_1(u_{m1}m_1^2 + u_{m2}m_2^2) + B_2 u_{m6} m_1 m_2 \\ &- B_1\left[\frac{B_1}{6c_{11}} + \frac{c_{12}}{c_{11}}(u_{m1} + u_{m2})\right]m_3^2 + \frac{1}{2}\mu_0 M_s^2(N_1 m_1^2 + N_2 m_2^2 \\ &+ N_3 m_3^2) - \mu_0 M_s(H_1 m_1 + H_2 m_2 + H_3 m_3), \qquad (2)\end{aligned}$$

where $m_i$ are the direction cosines of magnetization **M** with respect to the principal cubic axes, $K_1$ and $K_2$ are the anisotropy constants of fourth and sixth order at constant strains, $B_1$ and $B_2$ are the magnetoelastic coefficients, $c_{11}$, $c_{12}$, and $c_{44}$ are the elastic stiffnesses at constant magnetization, and $u_{m1}$, $u_{m2}$, $u_{m6}$ denote the in-plane film strains $u_{11}$, $u_{22}$, $u_{12}$ imposed by a dissimilar thick substrate. Differentiating Eq. (2), one readily finds the derivatives $\partial F/\partial m_i$ defining the sought effective field $\mathbf{H}_{\text{eff}}$.

Substituting $\mathbf{H}_{\text{eff}}$ into Eq. (1) and taking into account that in equilibrium $\mathbf{M} \times \mathbf{H}_{\text{eff}} = 0$, we derived a system of equations for the quantities $\delta m_i(t)$ describing the magnetization precession. Since $m_1^2 + m_2^2 + m_3^2 = 1$, one of these quantities can be expressed through two others. Focusing on the case of $m_3^0 \neq 0$, we used the relation $\delta m_3 = -(m_1^0 \delta m_1 + m_2^0 \delta m_2)/m_3^0$ and, neglecting the loss term in the first approximation, reduced the system of linearized equations to

$$\frac{M_s}{\gamma}\frac{d\delta m_1}{dt} + A_{11}\delta m_1 + A_{12}\delta m_2 = \mu_0 M_s m_3^0 h_2 - \mu_0 M_s m_2^0 h_3,$$
$$\frac{M_s}{\gamma}\frac{d\delta m_2}{dt} + A_{21}\delta m_1 + A_{22}\delta m_2 = -\mu_0 M_s m_3^0 h_1 + \mu_0 M_s m_1^0 h_3 \cdot \qquad (3)$$

Here $A_{ij}$ are polynomials of direction cosines $m_i^0$ of the static magnetization $\mathbf{M}_s$, in which coefficients depend on components $H_i$ of the static magnetic field, misfit strains $u_{m1}$, $u_{m2}$, $u_{m6}$, and the material parameters involved in Eq. (2). Since $A_{11} + A_{22} = 0$, the expression for the FMR frequency $\nu_r$ resulting from Eq. (3), which is valid to first order in the loss term,[26] takes the form

$$\nu_r = \frac{\gamma}{2\pi M_s}\sqrt{A_{11}A_{22} - A_{12}A_{21}} \cdot \qquad (4)$$

Using the relations $m_1^0 = \cos\varphi\sin\theta, m_2^0 = \sin\varphi\sin\theta$, and $m_3^0 = \cos\theta$, it is possible to rewrite Eq. (4) in terms of the orientation angles $\varphi$ and $\theta$ of the static magnetization $\mathbf{M}_s$. Indeed, these angles must correspond to the equilibrium orientation of $\mathbf{M}_s$ determined via minimization of the free energy given by Eq. (2). It should be noted that, at $\sin\theta \neq 0$, the FMR frequency can be also calculated from the formula $\nu_r = \gamma\sqrt{F_{\theta\theta}F_{\varphi\varphi} - F_{\theta\varphi}^2}/(2\pi M_s \sin\theta)$, where the quantities under the square root are the second derivatives of energy $F$ with respect to the angles $\theta$ and $\varphi$.[27] However, this formula cannot be used for the magnetization directions orthogonal to the film surfaces ($\sin\theta = 0$), which are important for our study. The procedure employed in Ref. 24, which involves the direct substitution of the strain-dependent effective magnetic field $\mathbf{H}_{\text{eff}}$ into the usual relation for the FMR frequency, does not provide rigorous theoretical description of the influence of lattice strains on the resonance conditions.

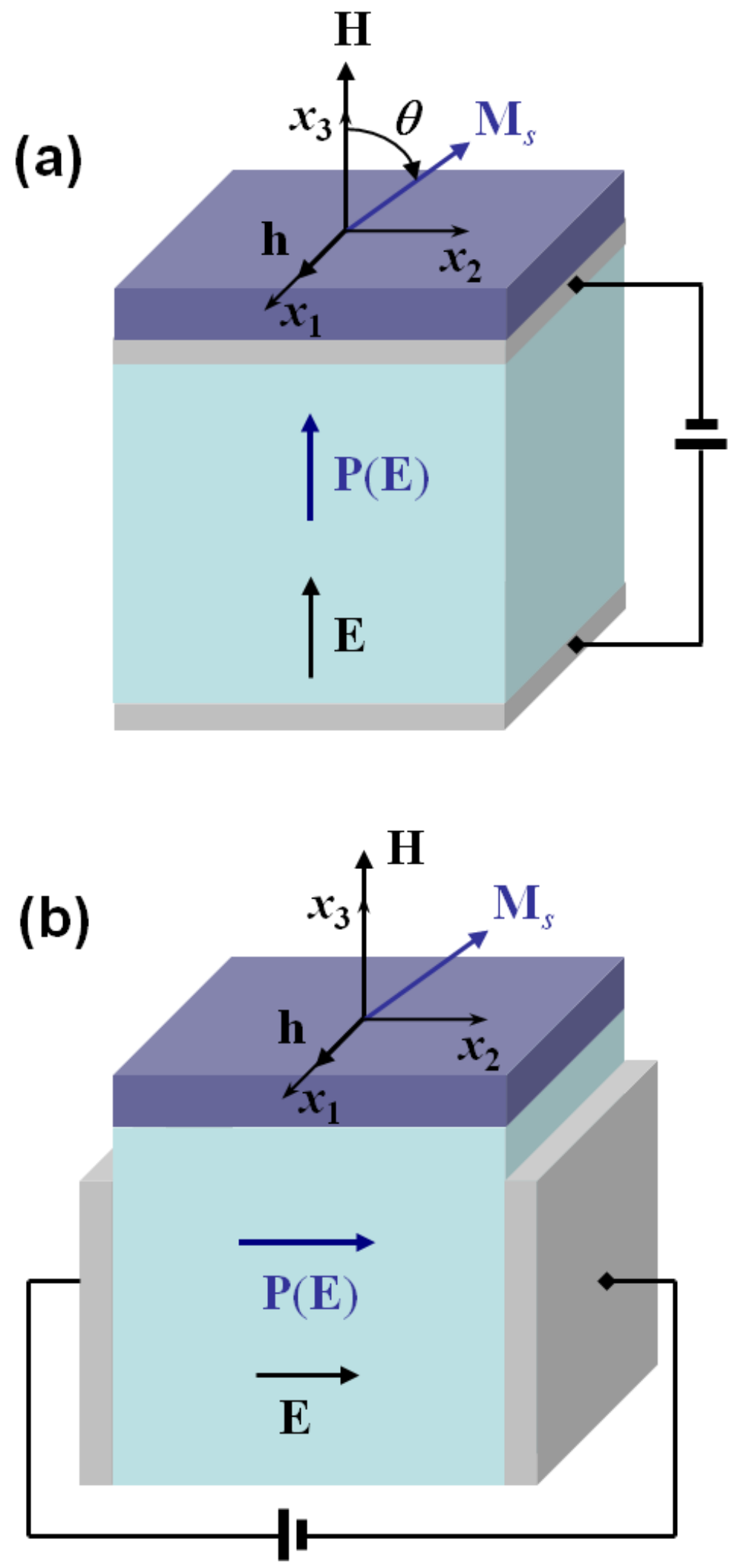

**FIG. 1.** Schematic drawing of a ferromagnetic film coupled to a ferroelectric substrate covered by two extended electrodes. A film with static magnetization $\mathbf{M}_s$ is subjected to the constant magnetic field **H** and the microwave magnetic field **h**. The polarization **P** of ferroelectric material varies with the applied electric field **E,** which creates macroscopic substrate deformations. Panel (a) shows the top-bottom arrangement of electrodes, whilst panel (b) describes their lateral configuration.

## III. STRAIN EFFECT ON FERROMAGNETIC RESONANCE

The theoretical analysis shows that the strain effect on FMR frequency has several remarkable features. Importantly, $\nu_r$ varies *linearly* only in the particular case of fully symmetric configuration, where the film with the out-of-plane magnetization $\mathbf{M}_s$ ($\theta = 0, \pi$) and equal in-plane demagnetizing factors ($N_1 = N_2$) is subjected to the orthogonal magnetic field ($H_1 = H_2 = 0$, $H_3 \neq 0$) and isotropic biaxial strain ($u_{m1} = u_{m2} = u_m$, $u_{m6} = 0$). For the resonance frequency $\nu_r$, under these conditions we obtain

$$\nu_r = \frac{\gamma}{2\pi}\left\{\mu_0 H_3 + \frac{1}{2}\mu_0 M_s(1-3N_3) + \frac{2}{M_s}\left[K_1^* + \frac{B_1^2}{6c_{11}} + \left(1+2\frac{c_{12}}{c_{11}}\right)B_1 u_m\right]\right\}, \tag{5}$$

where $K_1^* = K_1 + B_1^2/(2c_{11}) - B_2^2/(2c_{44})$. Equation (5) shows that both the FMR frequency $\nu_r$ and field $H_r$ linearly depend on $u_m$. It should be noted that $H_r$ increases with strain when $\nu_r$ decreases and vice versa.

As soon as the four-fold symmetry about the substrate normal is broken, the strain dependence of FMR frequency becomes *nonlinear*. In the particular case of $\varphi = \pi/2$, the variation of $\nu_r$ with lattice strains $u_{m1}$, $u_{m2}$, and $u_{m6}$ takes the form

$$\begin{aligned}\nu_r = \frac{\gamma}{2\pi M_s}\Bigg\{&\bigg[\mu_0 M_s H_3 \sec\theta + \mu_0 M_s^2(N_1 - N_3) + 2(K_1 + K_2\sin^2\theta)\cos^2\theta - \frac{2B_1^2}{3c_{11}}(1-3\cos^2\theta)\\ &- \frac{B_2^2}{c_{44}}\cos 2\theta + 2B_1\left(u_{m1} + \frac{c_{12}}{c_{11}}(u_{m1}+u_{m2})\right)\bigg]\bigg[\mu_0 M_s(H_2\sin\theta + H_3\cos\theta) + \mu_0 M_s^2(N_2 - N_3)\cos 2\theta\\ &+ 2K_1^*\cos 4\theta + 2B_1\left(\frac{B_1}{6c_{11}} + u_{m2} + \frac{c_{12}}{c_{11}}(u_{m1}+u_{m2})\right)\cos 2\theta\bigg] - B_2 u_{m6}(\mu_0 M_s H_1\sin\theta + B_2 u_{m6}\cos 2\theta)\Bigg\}^{1/2}.\end{aligned} \tag{6}$$

It can be seen that the deviation of $\theta$ from zero imparts nonlinearity even to the dependence of $\nu_r$ on the isotropic biaxial strain $u_m$. Using Eq. (6) together with Eq. (2), it is also possible to determine the strain dependence of the resonance magnetic field $H_r$ but this task is more involved as the orientation angle $\theta$ generally varies with the field intensity.

In turn, for a film with an in-plane static magnetization ($\theta = \pi/2$), the calculation gives

$$\begin{aligned}\nu_r = &\frac{\gamma}{2\pi M_s}\left[\mu_0 M_s H_\parallel + \mu_0 M_s^2(N_2 - N_1)\cos 2\varphi + 2K_1\cos 4\varphi + 2B_1(u_{m2} - u_{m1})\cos 2\varphi - 2B_2 u_{m6}\sin 2\varphi\right]^{1/2}\\ &\times\Bigg\{\mu_0 M_s H_\parallel + \mu_0 M_s^2(N_3 - N_2\sin^2\varphi - N_1\cos^2\varphi) + K_1^* + \frac{K_1}{2}(1+\cos 4\varphi) + \frac{K_2}{2}\sin^2 2\varphi + \frac{B_1^2}{6c_{11}}\\ &- 2B_1\left[u_{m1}\cos^2\varphi + u_{m2}\sin^2\varphi + \frac{c_{12}}{c_{11}}(u_{m1}+u_{m2})\right] - B_2 u_{m6}\sin 2\varphi\Bigg\}^{1/2},\end{aligned} \tag{7}$$

where $H_\parallel = H_1\cos\varphi + H_2\sin\varphi$ is the projection of $\mathbf{H}$ on the direction of static magnetization. Remarkably, at fixed orientation angle $\varphi$ and $u_{m1} = u_{m2} = u_m$, $u_{m6} = 0$, Eq. (7) reduces to a *square-root* dependence of resonance frequency on the isotropic biaxial strain $u_m$.

It should be emphasized that Eq. (5) and other analytic relations for the strain dependence of FMR frequency hold only for certain ranges of misfit strains, within which the equilibrium direction of magnetization remains fixed. In general, the orientation of $\mathbf{M}_s$ depends on misfit strains as well,[22] which affects the FMR parameters additionally. To study the influence of strain-induced magnetization reorientations on resonance conditions, we performed numerical calculations for epitaxial films of $Fe_{60}Co_{40}$ and $CoFe_2O_4$.[28] Figure 2(a) shows the direction cosines $m_i$ of magnetization as a function of biaxial strain $u_m$ calculated for a $Fe_{60}Co_{40}$ film subjected to an out-of-plane magnetic field $H_3$ = 25 kOe. It can be seen that the magnetization $\mathbf{M}_s$ is orthogonal to the film surfaces only at strains below $u_m^* \cong 0.39\%$. Above this critical strain, $\mathbf{M}_s$ starts to rotate gradually towards the in-plane [010] direction, which is accompanied by a *qualitative change* in the strain dependence of FMR frequency $\nu_r$. As demonstrated by Fig. 2(b), the linear decrease of $\nu_r$ with strain is replaced above $u_m^*$ by a nonlinear increase. Remarkably, the FMR frequency goes to zero at $u_m = u_m^*$, where the *second-order* SRT takes place.

In contrast, the strain-induced SRT in a $CoFe_2O_4$ film subjected to the magnetic field $H_3$ = 5 kOe is of the *first order*. This feature is evidenced by an abrupt reorientation of magnetization at $u_m^* \cong -0.06\%$ demonstrated by Fig. 3(a). As a result, the FMR frequency remains finite at the critical strain $u_m^*$, but experiences a step-like drop across the SRT [see Fig. 3(b)]. It can be also seen that the resonance frequency is very sensitive to the tensile strains in an epitaxial $CoFe_2O_4$ film. Remarkably, the FMR can be shifted to ultrahigh frequencies exceeding 200 GHz by a biaxial strain of less than 1%.

Suppose now that the ferromagnetic film is mechanically coupled to a thick ferroelectric substrate, as shown in Fig. 1. In this case, the film strains $u_{ij}$ can be changed by an electric field $\mathbf{E}$ applied to the substrate.[22] Indeed, owing to the converse piezoelectric effect inherent in ferroelectric materials, the applied field creates macroscopic deformations $u_{ij}^s = d_{kij} E_k$ in the substrate with piezoelectric coefficients $d_{kij}$. The deformations $u_{\alpha\beta}^s$ ($\alpha$, $\beta$ =1, 2) appearing in the (001) planes parallel to the film/substrate interface are transmitted to the film via the interfacial bonding. In the case of perfect transmission expected for epitaxial heterostructures, the in-plane film strains become $u_{\alpha\beta}(\mathbf{E}) = u_{\alpha\beta}^0 + u_{\alpha\beta}^s(\mathbf{E})$, where $u_{\alpha\beta}^0$ denote their initial values defined by the misfit strains $u_{m1}$, $u_{m2}$, and $u_{m6}$.

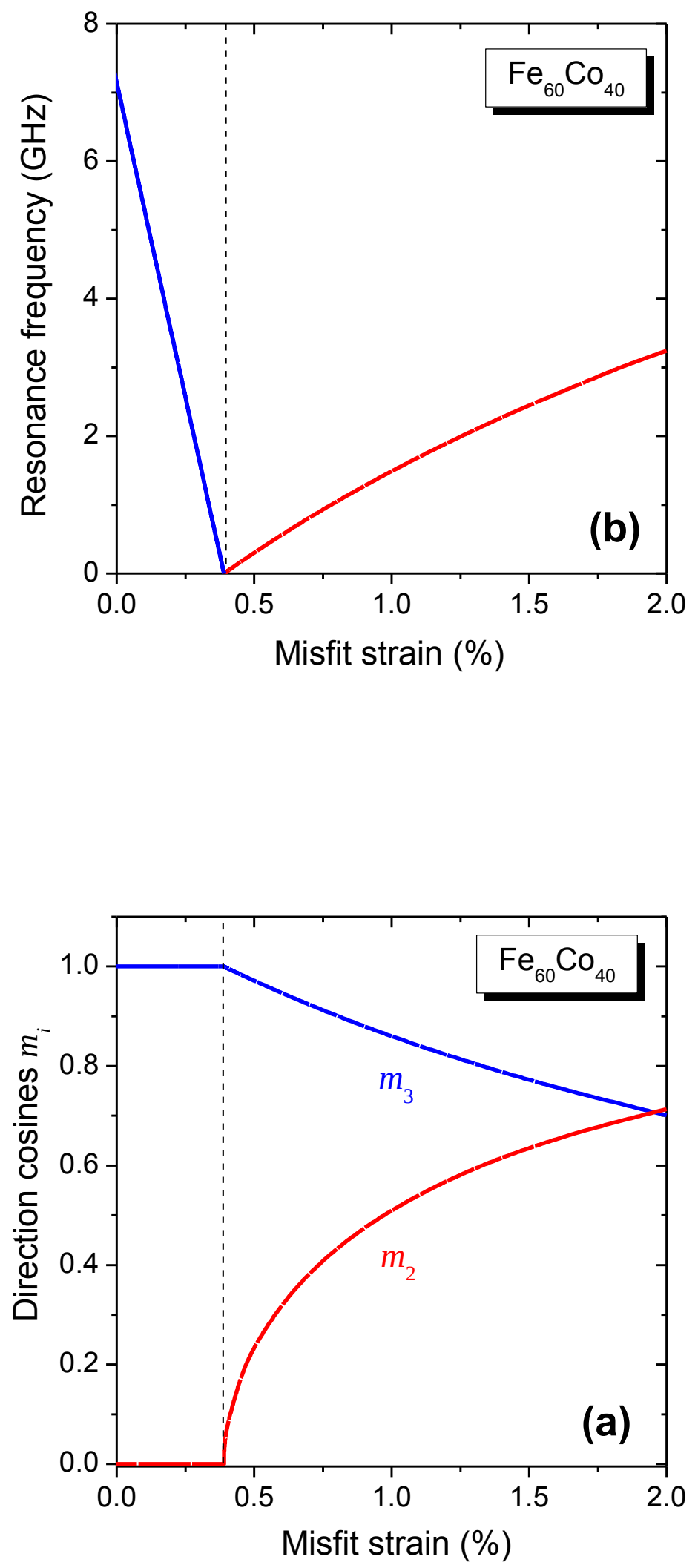

**FIG. 2.** Magnetization orientation and FMR frequency calculated for strained $Fe_{60}Co_{40}$ epitaxial films. The direction cosines $m_i$ of the equilibrium magnetization $\mathbf{M}_s$ (a) and the FMR frequency $\nu_r$ (b) are shown as a function of the misfit strain in the film/substrate system. The film is assumed to be under the static magnetic field $H_3$ = 25 kOe orthogonal to the film surfaces. The dashed line indicates the strain-induced spin reorientation transition.

When the electric field is orthogonal to the interface and the film/substrate system has the four-fold symmetry about the substrate normal, we obtain $u_{11}(E_3) = u_{22}(E_3) = u_m + d_{31}E_3$ and $u_{12}(E_3) = 0$ since $u_{m1}$ = $u_{m2}$ = $u_m$, $u_{m6}$ = 0, $d_{311}$ = $d_{322}$ = $d_{31}$, and $d_{312}$ = 0. Accordingly, the influence of electric field $E_3$ on the FMR frequency $\nu_r$ becomes similar to that of the isotropic strain $u_m$. In particular, for the film with out-of-plane static magnetization, from Eq. (5) we find that the tunability $\partial\nu_r/\partial E_3$ of FMR frequency is independent of the misfit strain in the heterostructure and the magnetic field $H_3$ and equals

$$\frac{\partial \nu_r}{\partial E_3} = \frac{\gamma}{\pi M_s}\left(1 + 2\frac{c_{12}}{c_{11}}\right)B_1 d_{31}. \tag{8}$$

Taking the relaxor ferroelectric $Pb(Zn_{1/3}Nb_{2/3})O_3$–4.5%$PbTiO_3$ (PZN-4.5%PT) as a representative substrate material, with $d_{31}$ = −1000 pm/V (Ref. 29) we calculate the tunability to be about 0.18 GHz/(kV/cm) for $Fe_{60}Co_{40}$, −2.05 GHz/(kV/cm) for $CoFe_2O_4$, and −0.23 GHz/(kV/cm) for Ni.[30] When the film magnetization deviates from the [001] direction, the magnitude of $\partial\nu_r/\partial E_3$ becomes smaller than in the out-of-plane magnetic state. As follows from Figs. 2(b) and 3(b), the tunability changes sign at the strain-induced SRT and starts to vary with the misfit strain $u_m$.

Instead of the top-bottom electrode arrangement, one can also cover the ferroelectric substrate by lateral electrodes (see Fig. 1). Since in this configuration the electric field is parallel to the interface, the four-fold symmetry about the substrate normal becomes broken even in heterostructures with $u_{m1}$ = $u_{m2}$ and $u_{m6}$ = 0. As a result, the influence of in-plane electric field on FMR parameters appears to be very different from that of the isotropic biaxial strain $u_m$. In particular, when the field direction is parallel to the [100] crystallographic axis of the film, the in-plane lattice strains become $u_{11}(E_1) = u_m + d^*_{33}E_1$ and $u_{22}(E_1) = u_m + d^*_{31}E_1$, where $d^*_{in}$ are the substrate piezoelectric coefficients defined in the rotated reference frame $(x^*_1, x^*_2, x^*_3)$ with the $x^*_3$ axis

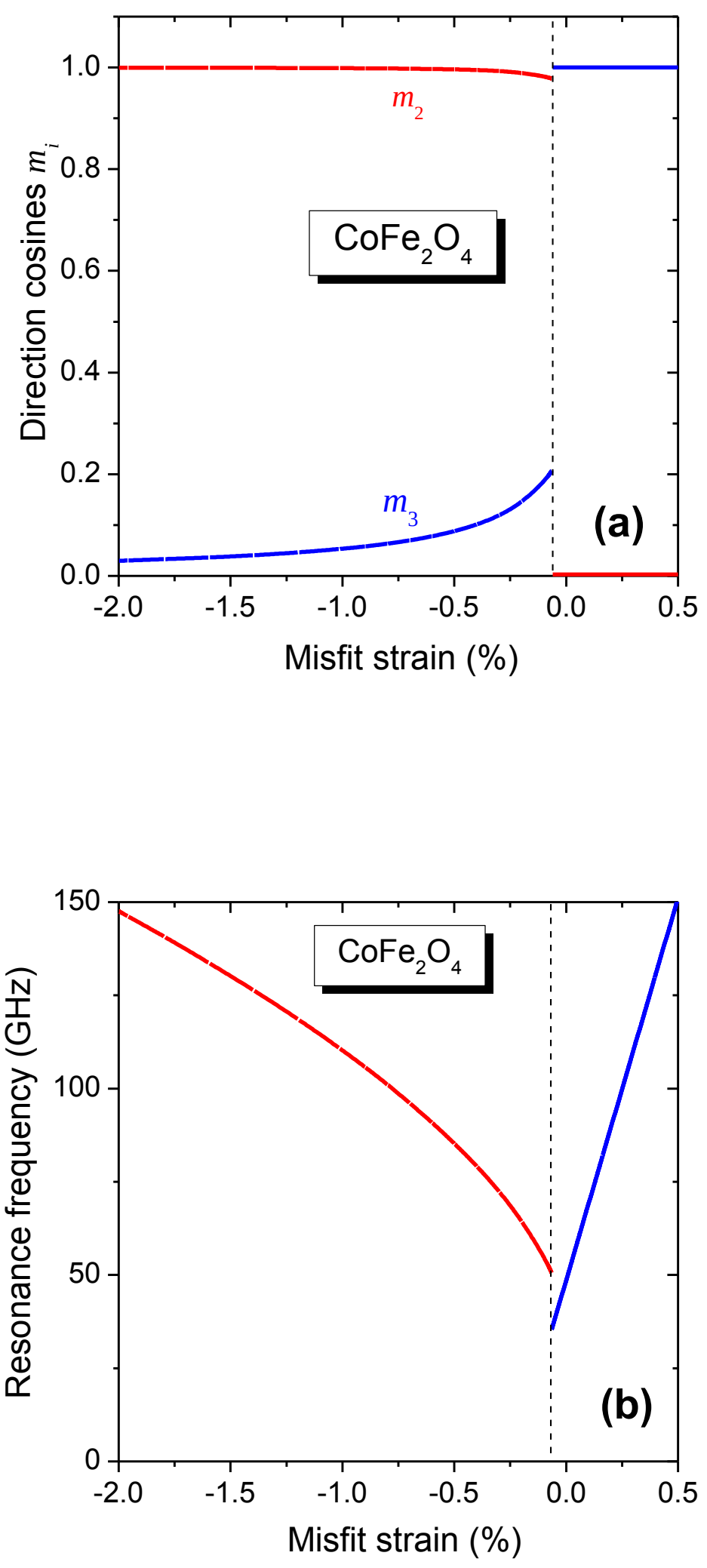

**FIG. 3.** Magnetization orientation and FMR frequency calculated for strained $CoFe_2O_4$ epitaxial films. The direction cosines $m_i$ of the spontaneous magnetization $\mathbf{M}_s$ (a) and the FMR frequency $\nu_r$ (b) are shown as a function of the misfit strain in the film/substrate system. The film is assumed to be under the static magnetic field $H_3$ = 5 kOe orthogonal to the film surfaces. The dashed line indicates the strain-induced spin reorientation transition.

oriented along the field direction and the $x^*_1$ axis parallel to the interface. Substituting these relations into Eq. (6), we obtain the following analytic formula for the FMR frequency of a film with the out-of-plane magnetization subjected to vertical magnetic field:

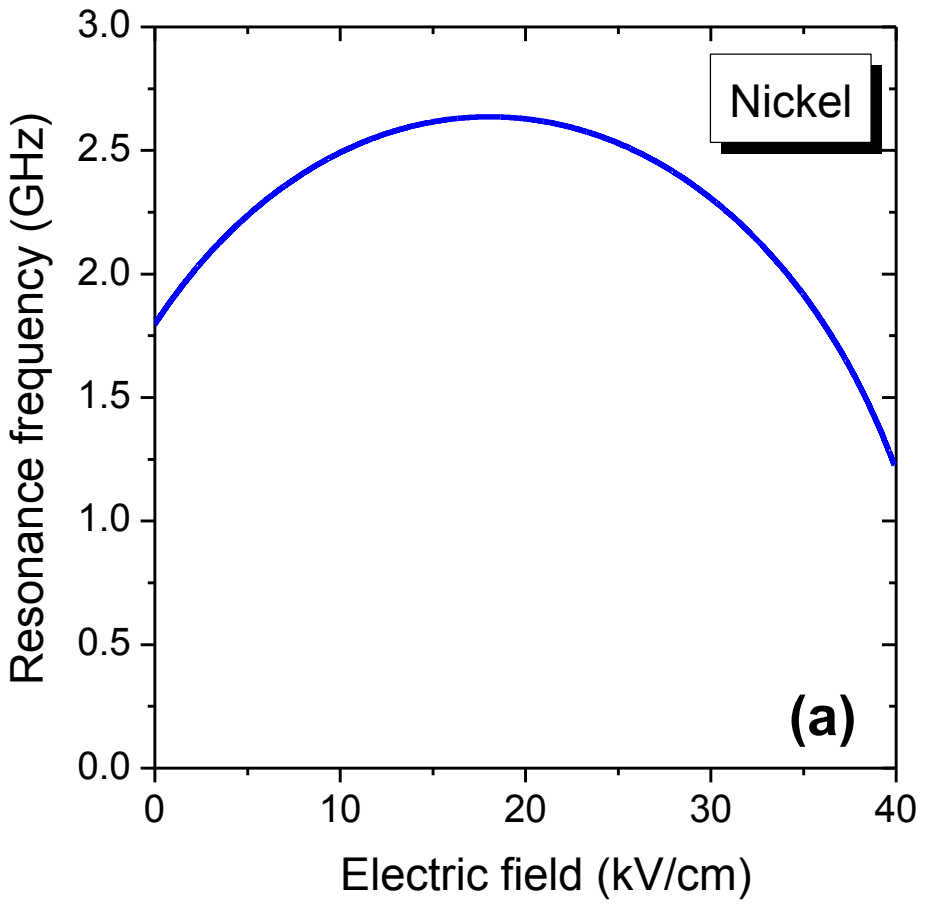

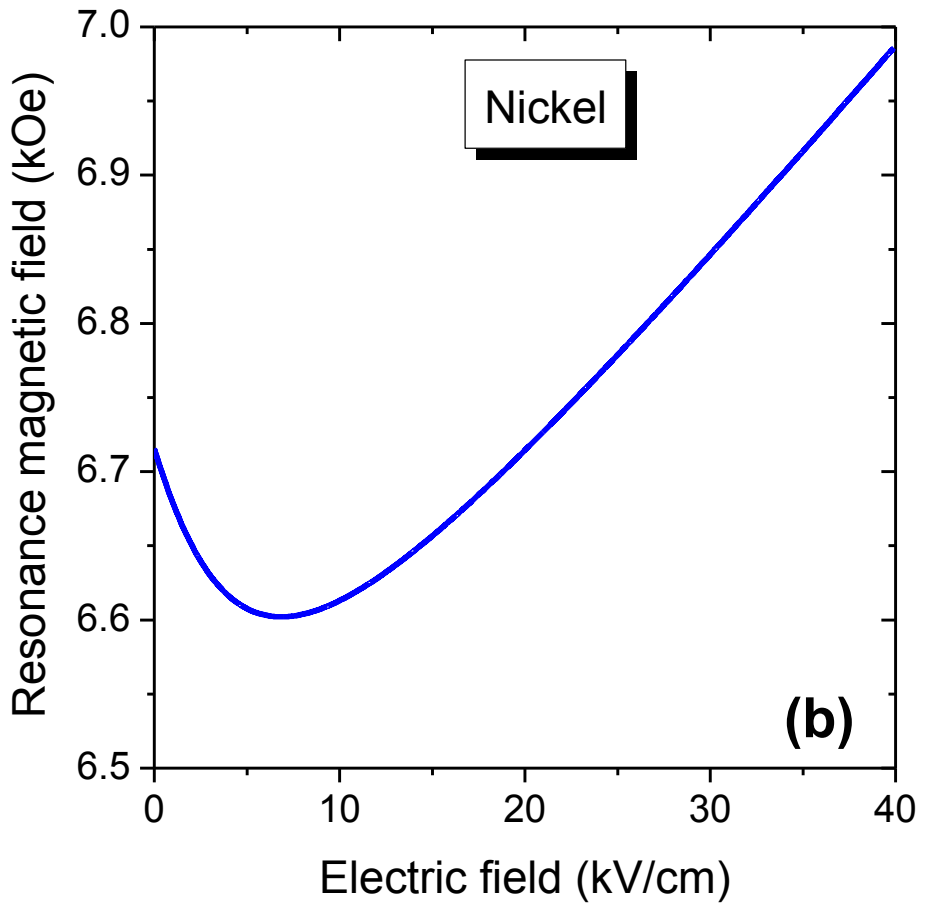

**FIG. 4.** FMR frequency $\nu_r$ (a) and field $H_r$ (b) of an epitaxial Ni film at zero initial misfit strain ($u_m$ = 0). The electric field **E** applied to the PZN-4.5%PT substrate is parallel to the interface, being directed along the [100] crystallographic axis of the film. The FMR frequency is given for the static magnetic field $H_3$ = 7 kOe orthogonal to the film surfaces, and the critical magnetic field corresponds to the frequency $\nu$ = 1 GHz of the microwave field.

$$\nu_r = \frac{\gamma\mu_0}{2\pi}\left\{H_3 + M_s(N_1 - N_3) + \frac{2}{\mu_0 M_s}\left[K_1^* + \frac{B_1^2}{6c_{11}} + \left(1 + 2\frac{c_{12}}{c_{11}}\right)B_1 u_m + B_1\left(d_{33}^* + \frac{c_{12}}{c_{11}}(d_{33}^* + d_{31}^*)\right)E_1\right]\right\}^{1/2}$$
$$\left\{H_3 + M_s(N_2 - N_3) + \frac{2}{\mu_0 M_s}\left[K_1^* + \frac{B_1^2}{6c_{11}} + \left(1 + 2\frac{c_{12}}{c_{11}}\right)B_1 u_m + B_1\left(d_{31}^* + \frac{c_{12}}{c_{11}}(d_{33}^* + d_{31}^*)\right)E_1\right]\right\}^{1/2}. \quad (9)$$

Since the piezoelectric coefficients involved in Eq. (9) are different in sign and $|d_{31}^*| \sim d_{33}^*/2$, the field dependence of FMR frequency has the form $\nu_r \sim \sqrt{(1 + pE_1)(1 - qE_1)}$, where parameters $p$ and $q$ are positive. Hence the resonance frequency may vary with electric field *nonmonotonically* even at a fixed orientation of the static magnetization. This remarkable feature is illustrated in Fig. 4(a), where the dependence $\nu_r(E_1)$ is plotted for the initially unstrained ($u_m$ = 0) Ni film under the magnetic field $H_3$ = 7 kOe.[31] The critical magnetic field $H_r(E_1)$ providing FMR in such a film also varies nonmonotonically, as shown in Fig. 4(b). Since the magnetization remains orthogonal to the film surfaces, the dependence $H_r(E_1)$ at constant frequency $\nu$ of microwave field can be described by an analytic relation

$$H_r = \sqrt{\left(\frac{2\pi\nu}{\mu_0\gamma}\right)^2 + \left[\frac{1}{2}M_s(N_1 - N_2) + \frac{B_1}{\mu_0 M_s}(d_{33}^* - d_{31}^*)E_1\right]^2} + \frac{1}{2}M_s(2N_3 - N_1 - N_2)$$
$$- \frac{2}{\mu_0 M_s}\left(K_1^* + \frac{B_1^2}{6c_{11}}\right) - \frac{B_1}{\mu_0 M_s}\left(1 + 2\frac{c_{12}}{c_{11}}\right)\left[2u_m + (d_{33}^* + d_{31}^*)E_1\right]. \quad (10)$$

It can be seen that nonlinearity of $H_r(E_1)$ results from the strain anisotropy created by in-plane electric field.

The electric field may also affect the equilibrium magnetization orientation in an epitaxial film.[22] In this situation, the field dependence of FMR parameters may be very different from the case of fixed magnetization orientation. Figure 5 shows the direction cosines $m_i(E_1)$ of static magnetization and the resonance frequency $\nu_r(E_1)$ calculated for the Ni film strained initially by –0.07%. At this value of the misfit strain $u_m$, the magnetization remains orthogonal to the film surfaces up to $E_1^* \cong$ 4.55 kV/cm and rotates gradually towards the in-plane [010] orientation at field intensities $E_1 > E_1^*$. As is seen from Fig. 5(b), the electric field dependence of $\nu_r$ changes dramatically across the SRT, which occurs at the critical intensity $E_1^*$. Remarkably, the FMR frequency goes to zero at this second-order phase transition. Moreover, the tunability $\partial\nu_r/\partial E_1$ of resonance frequency has an anomaly at the SRT, diverging at $E = E_1^*$.

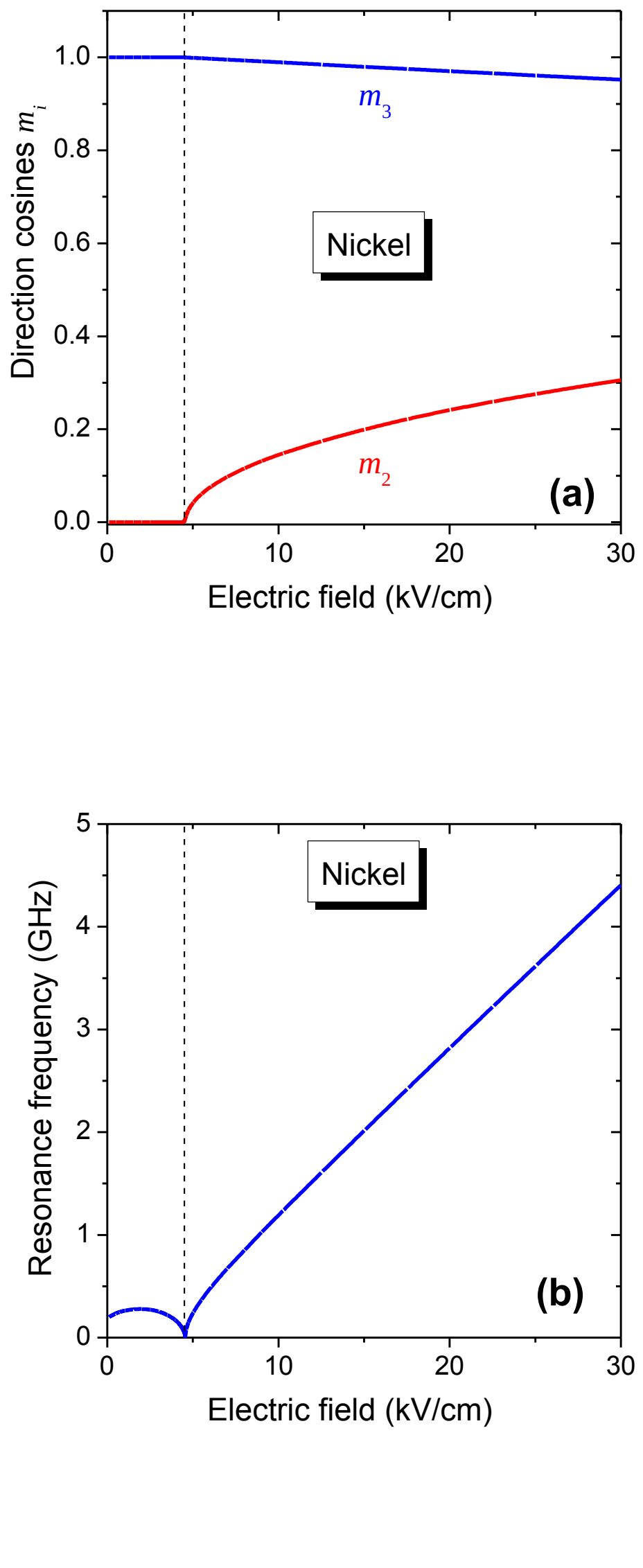

**FIG. 5.** Direction cosines $m_i$ of the magnetization $\mathbf{M}_s$ (a) and the FMR frequency $\nu_r$ (b) of a Ni film plotted as a function of the electric field applied to the PZN-4.5%PT substrate along the interface. The initial misfit strain $u_m$ is taken to be –0.07%. The film is assumed to be under the static magnetic field $H_3$ = 7 kOe orthogonal to the film surfaces. The dashed line indicates the strain-induced SRT.

## IV. CONCLUSIONS

Thus, we have shown that substrate-induced lattice strains represent an efficient tool for tuning the FMR parameters of ferromagnetic films. Our main theoretical predictions can be summarized as follows:

(i) A linear variation of the FMR frequency $\nu_r$ appears only in the symmetric case, where the static magnetic field **H** and magnetization $\mathbf{M}_s$ are orthogonal to the surfaces of a film subjected to isotropic biaxial in-plane strain $u_m$.

(ii) A *nonmonotonic* variation of the resonance frequency $\nu_r$ with the isotropic strain $u_m$ is revealed across the strain-induced spin reorientation transitions (SRTs). Remarkably, $\nu_r$ reaches minimum at a critical strain $u_m^*$ at which the SRT takes place.

(iii) The FMR frequency of a ferromagnetic film coupled to a ferroelectric substrate may change dramatically under the influence of electric field **E** applied to the substrate. The tunability $\partial\nu_r/\partial E$ of resonance frequency strongly depends on the field intensity $E$, displaying a change in sign across the field-induced SRT and a drastic increase in magnitude near the critical field.

(iv) The resonance magnetic field $H_r$ measured at a constant microwave frequency may also vary nonmonotonically with the electric field applied to the ferroelectric substrate.

These predictions provide guidelines for the fabrication of advanced linear and nonlinear microwave devices with improved performances. In particular, owing to strongly enhanced electric tunability of FMR frequency, properly strained thin-film hybrids should be superior to bulk ferromagnetic-ferroelectric composites for applications in electrically tunable microwave devices with high tuning speed and low energy consumption. Moreover, the tunability itself can be varied in a controlled manner by a bias voltage applied to a ferroelectric substrate. In the linear regime, multiferroic hybrids can be used not only for magnetoelectric bandpass filters and phase shifters,[32,33] but also for resonant circuits in tunable oscillators with high tuning speed and for frequency tunable resonant antennas. If the ferromagnetic films are exploited in the nonlinear regime, they can be applied in parametric circuits, mixers, and frequency multipliers. These components can represent essential building blocks of high speed reconfigurable microwave systems, such as radar and communication systems or phased antenna arrays. We hope that our theoretical results will trigger experimental studies of the strain and voltage effects on the FMR in ferromagnetic thin films and nanostructures and the development of novel magnetoelectric devices. Finally, it should be noted that our approach also opens the way for nonlinear thermodynamic calculations of FMR parameters of multiferroic columnar nanostructures.[34,35]

## ACKNOWLEDGMENTS

This work was financially supported by the Deutsche Forschungsgemeinschaft via the DFG grant INST 257/343-1/570236 (Mercator Visiting Professorship awarded to N. A. P.) and the SFB 855 01/10. The authors also thank Franz Faupel and Michael Hambe for critical reading of the manuscript.